# Constraints on the Coupling between Axionlike Dark Matter and Photons Using an Antiproton Superconducting Tuned Detection Circuit in a Cryogenic Penning Trap

Jack A. Devlin[1,2,*], Matthias J. Borchert,[1,3,4] Stefan Erlewein,[1,2] Markus Fleck,[1,5] James A. Harrington,[1,6] Barbara Latacz,[1] Jan Warncke,[1] Elise Wursten,[1,2] Matthew A. Bohman,[1,6] Andreas H. Mooser,[1,6] Christian Smorra,[1,7] Markus Wiesinger,[1,6] Christian Will,[6] Klaus Blaum,[6] Yasuyuki Matsuda,[5] Christian Ospelkaus,[3,4] Wolfgang Quint,[8] Jochen Walz,[7,9] Yasunori Yamazaki,[1] and Stefan Ulmer[1]

[1]*RIKEN, Ulmer Fundamental Symmetries Laboratory, 2-1 Hirosawa, Wako, Saitama 351-0198, Japan*
[2]*CERN, Esplanade des Particules 1, CH-1211 Geneva 23, Switzerland*
[3]*Physikalisch-Technische Bundesanstalt, Bundesallee 100, D-38116 Braunschweig, Germany*
[4]*Institut für Quantenoptik, Leibniz Universität Hannover, Welfengarten 1, D-30167 Hannover, Germany*
[5]*Graduate School of Arts and Sciences, University of Tokyo, 3-8-1 Komaba, Tokyo 153-8902, Japan*
[6]*Max-Planck-Institut für Kernphysik, Saupfercheckweg 1, D-69117 Heidelberg, Germany*
[7]*Institut für Physik, Johannes Gutenberg-Universität, Staudinger Weg 7, D-55128 Mainz, Germany*
[8]*GSI-Helmholtzzentrum für Schwerionenforschung GmbH, Planckstraße 1, D-64291 Darmstadt, Germany*
[9]*Helmholtz-Institut Mainz, Johannes Gutenberg-Universität, Staudinger Weg 18, D-55128 Mainz, Germany*

We constrain the coupling between axionlike particles (ALPs) and photons, measured with the superconducting resonant detection circuit of a cryogenic Penning trap. By searching the noise spectrum of our fixed-frequency resonant circuit for peaks caused by dark matter ALPs converting into photons in the strong magnetic field of the Penning-trap magnet, we are able to constrain the coupling of ALPs with masses around 2.7906–2.7914 $neV/c^2$ to $g_{a\gamma} < 1 \times 10^{-11}$ $GeV^{-1}$. This is more than one order of magnitude lower than the best laboratory haloscope and approximately 5 times lower than the CERN axion solar telescope (CAST), setting limits in a mass and coupling range which is not constrained by astrophysical observations. Our approach can be extended to many other Penning-trap experiments and has the potential to provide broad limits in the low ALP mass range.

DOI: 10.1103/PhysRevLett.126.041301

Quantum chromodynamic (QCD) axions are hypothetical particles that would explain the apparent conservation of charge-parity symmetry by the strong force [1]. In addition to the QCD axion, extensions to the standard model also predict several new axionlike particles (ALPs), with a range of possible masses and coupling constants [2,3]. These particles are excellent dark matter candidates, since they would be produced in the early Universe and form a cold dark matter halo consistent with astrophysical observations [4]. A number of laboratory experiments and astrophysical observations have placed limits on ALP masses and couplings in the $neV/c^2$ range [5,6]. Intriguingly, some astrophysical analyses have suggested that ALPs with masses of a few $neV/c^2$ could explain higher-than-expected $\gamma$-ray transparency [7,8] and energy-dependent modulations in the $\gamma$-ray spectra of certain pulsars [9], although these hints are in a region disfavored by theoretical models which suggest ALPs are dark matter [4]. Nevertheless, these astrophysical indications make it highly desirable to study the coupling between potential low-mass ALPs and photons directly with laboratory experiments.

ALPs couple to **E** and **B** fields through the Lagrange density term

$$\mathcal{L}_{a\gamma\gamma} = -g_{a\gamma} a(\mathbf{x})\mathbf{E}(\mathbf{x}) \cdot \mathbf{B}(\mathbf{x}), \quad (1)$$

where $a(\mathbf{x})$ is the local ALP field and $g_{a\gamma}$ is a coupling constant [5,10]. For the QCD axion there is an inverse relationship between $g_{a\gamma}$ and the axion mass $m_a$, but for an ALP the two quantities are independent. Any dark matter ALPs would form a classical field oscillating with a characteristic frequency close to the ALP Compton frequency $m_a c^2/h$. Equation (1) allows ALPs to convert into photons in a strong magnetic field. To detect the conversion of dark matter ALPs with very low masses, Sikivie and collaborators proposed an extension of the haloscope concept [11] using sensitive resonant $LC$ circuits in strong magnetic fields [12] to measure the weak magnetic field oscillations sourced by ALPs. A number of current and proposed experiments are seeking to detect ALPs in this mass range [13–20]. In this work we show how, depending on the coil orientation, an ultrasensitive superconducting single-particle detector of a cryogenic Penning-trap experiment [21,22] can also detect ALPs. Although these devices are not dedicated axion detectors, they are able to set strong nonastrophysical limits on the existence of ALPs in a narrow band around their resonance frequency. By combining detector data from many Penning-trap experiments, it should be possible to search for ALP signals over a significant range of frequencies. In this analysis, we set limits on the ALP-to-photon coupling strength using the axial detection system of the analysis trap of the baryon antibaryon symmetry experiment (BASE) [23], complementing our study of the possible interactions between ALPs and antiprotons [24].

BASE is a cryogenic Penning-trap experiment located at CERN's Antimatter Factory [23], dedicated to testing charge-parity-time-reversal invariance by comparing the fundamental properties of protons and antiprotons [25,26]. An illustration of the analysis trap (AT) which is used to determine the antiproton spin state in high-precision magnetic moment measurements [26,27] is shown in Fig. 1(a). It comprises a stack of cylindrical gold-plated copper and Co/Fe ring electrodes, which are separated by sapphire spacers, and placed into the 1.945 T axial magnetic field of a horizontal superconducting magnet. An antiproton is confined radially by the magnetic field and axially by voltages applied to the electrodes. As the particle oscillates, femtoamp-sized image currents are induced in the trap electrodes, which are picked up using high-sensitivity $LC$ circuits as image-current detectors.

The $LC$ circuit [21], which is used both to detect antiproton image currents and to extract the ALP-photon interaction limits presented in this work, is formed by connecting one end of a toroidal superconducting inductor, shown on the right-hand side of Fig. 1(a), to an electrode, while the other end is grounded. The remaining electrodes are low-pass filtered so that they are held at radio-frequency ground. The inductor is composed of $N_T \simeq 1100$ turns of 120-$\mu$m-diameter superconducting wire wound around a cylindrical polytetrafluoroethylene (PTFE) former of inner radius $r_1 = 11.5$ mm, outer radius $r_2 = 19$ mm, and length $l = 22$ mm. The inductor is placed inside a NbTi cylindrical housing. A wire tap is connected to couple the inductor to the amplifier chain [21], defining the amplifier coupling factor $\kappa \simeq 0.2$. The magnetic field is $\|\mathbf{B}_e\| = 1.85(5)$ T at the position of the AT detector, directed along the axis of the toroid, as indicated by the red arrow in Fig. 1(a). Figure 1(b) shows an effective circuit diagram of the particle detector. The inductor and the parasitic capacitance $C_p$ of the trap electrode form an $LC$ circuit with a $Q$ factor of $4.2(3) \times 10^4$, resonance frequency $\nu_z \simeq 674.9$ kHz, and effective parallel resistance $R_p \simeq 2\pi Q \nu_z L \simeq 288$ M$\Omega$, where $L$ is the inductance of the circuit. When an antiproton reaches thermal equilibrium with the detector, it acts like a series $LC$ circuit, shown in blue in Fig. 1(b). By adjusting the voltages applied to the trap electrodes, the particle's axial oscillation frequency can be tuned to resonance with the detector, leading to a voltage drop $V_p = R_p I_p$ across the resonator.

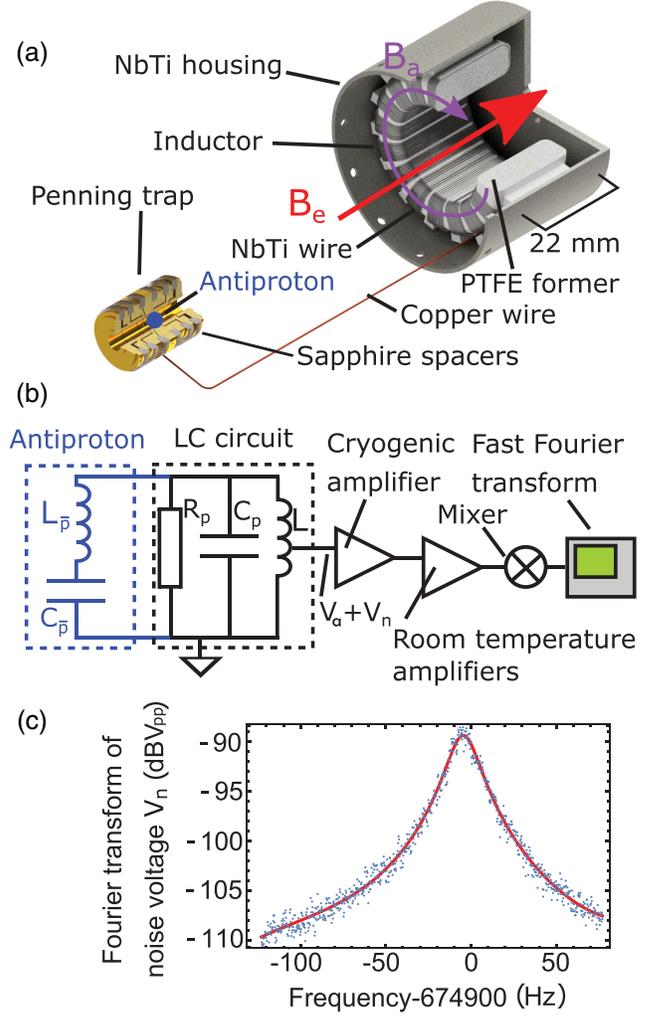

FIG. 1. (a) An illustration of the main elements of the cryogenic detection system together with the external magnetic field $B_e$ and the azimuthal ALP magnetic field $B_a$. The NbTi end cap is not shown for clarity. (b) The effective circuit diagram for the detection system. When an antiproton is in thermal equilibrium with the detector, as is the case during temperature measurements, the trapped particle behaves like the series $LC$ circuit shown in blue. During ALP searches, the particle's axial frequency is out of resonance with the detector, so the blue part of the circuit can be ignored. (c) A single Fourier transformed spectrum of the voltage noise $V_n$ recorded with 60 s averaging. The red line plots Eq. (6) with parameter values $\hat{\mathbf{b}}$ found by maximizing $\mathcal{L}(d|\{g_{a\gamma} = 0, \hat{\mathbf{b}}\})$ for this dataset $d$.

As well as being ideally suited to detecting single-particle image currents, the resonant $LC$ circuit is also sensitive to changes in the magnetic flux within the toroidal inductor, achieving maximum sensitivity when the frequency of those flux changes matches the resonant frequency of the LC circuit. The oscillating ALP field can source oscillating magnetic fields which produce such flux changes [12]. To calculate the strength of these fields, we note that the presence of an ALP field $a$ modifies Maxwell's equations; in particular, in the presence of a strong external magnetic field $\mathbf{B}_e$ and no free electric current density, the Maxwell-Ampere equation becomes

$$\nabla \times \mathbf{B} - \mu_0 \dot{\mathbf{D}} = -g_{a\gamma} \mathbf{B}_e \dot{a}, \quad (2)$$

where $\mathbf{B}$ and $\mathbf{D}$ are the usual classical fields. By solving Eq. (2) for the boundary conditions of the superconducting resonator [28,29], noting that $\mathbf{B}_e$ points along the resonator's axis of symmetry and its dimensions are much smaller than the ALP wavelength $\lambda_a = h/(m_a c)$, we find that the right-hand-side term sources an azimuthal magnetic field $\mathbf{B}_a$, shown in purple in Fig. 1(a), which oscillates at a frequency $\nu_a = (m_a/2h)(2c^2 + \mathbf{v} \cdot \mathbf{v})$, where $\mathbf{v}$ is the ALP velocity and the rms amplitude is

$$\mathbf{B}_a = -\frac{1}{2} g_{a\gamma} r \sqrt{\rho_a \hbar c} \|\mathbf{B}_e\| \hat{\boldsymbol{\phi}}. \quad (3)$$

Here, $\rho_a \hbar c = 2\pi^2 \nu_a^2 |a|^2$ is the local ALP energy density, $r$ is the radial distance from the axis of the toroid, and $\hat{\boldsymbol{\phi}}$ is a unit vector in the azimuthal direction. As well as the magnetic field given by Eq. (3), there is also a heavily suppressed axial electric field, which we cannot detect in this apparatus. Equivalent electric and magnetic fields are also generated inside the Penning trap; however, the effect of these fields on both the motion of the antiproton and the detection circuit [30] is negligible in comparison to the effect of the magnetic field in the detection circuit given by Eq. (3) which we consider in this work.

Because of the orientation of the toroidal coil, the oscillating magnetic field leads to a changing flux in the inductor, which, in turn, produces an oscillating voltage at the input of the first cryogenic amplification stage indicated in Fig. 1(b), whose rms amplitude is given by

$$V_a = \frac{\pi}{2} Q \sqrt{f(\nu, Q, \mathbf{q})} \kappa \nu_a l N_T (r_2^2 - r_1^2) g_{a\gamma} \|\mathbf{B}_e\| \sqrt{\rho_a \hbar c}. \quad (4)$$

The function $\sqrt{f(\nu, Q, \mathbf{q})} = |Z(\nu)|/R_p$ is the frequency dependence of the resonator's impedance $Z(\nu)$ divided by its value on resonance at $\nu = \nu_0$, where $|Z(\nu_0)| = R_p = 288$ MΩ. The function $f(\nu, Q, \mathbf{q})$ can be approximated by

$$f(\nu, Q, \mathbf{q}) = \frac{1}{1 + \frac{4Q^2(\nu-\nu_0)^2}{\nu_0^2}} + h(\nu, \mathbf{q}), \quad (5)$$

where the first term is a Lorentzian centered on $\nu_0$ and $h(\nu, \mathbf{q})$ are correction functions which account for tiny spectral asymmetries and depend on free phenomenological parameters $\mathbf{q}$. In addition to the possible ALP signal, there is also Johnson noise from the impedance of the LC resonator at a temperature $T_z$ and the amplifier's equivalent input noise $e_n$, which contribute an rms voltage noise at the input of the amplifier of

$$V_n = \sqrt{e_n^2 \Delta\nu + \kappa^2 4 k_B T_z \Delta\nu R_p f(\nu, Q, \mathbf{q})}, \quad (6)$$

where $\Delta\nu$ is the width of a single spectrum analyzer bin. Both voltages $V_a$ and $V_n$ are amplified [21], and the resulting voltage signal is recorded by an audio analyzer [31], which performs a fast Fourier transform (FFT). A plot of the FFT is shown in Fig. 1(c).

Uniquely, we are able to use a single trapped antiproton whose axial motion is in equilibrium with the detection system to measure the parallel resistance $R_p$ and the noise temperature $T_z$ of the circuit. This enables the measured voltage noise signal to be directly related to an expected ALP-to-photon conversion limit. To measure the parallel resistance $R_p$, first the trapping voltage is adjusted so that the axial oscillation frequency of the antiproton matches the resonant frequency of the detector. The particle behaves like an effective series LC circuit [shown in blue in Fig. 1(b)] and shorts the resonator noise, leading to a sharp dip with a full width at half maximum $\delta\nu = 5.6$ Hz, from which the parallel resistance $R_p = 2\pi(D_{\text{eff}}/q)^2 \delta\nu m_{\bar{p}}$ can be found [32]. Here, $D_{\text{eff}} = 11.2$ mm is a trap-specific length, $q$ is the antiproton charge, and $m_{\bar{p}}$ is the antiproton mass. For a resonant LC circuit, $R_p$ and $Q$ are simply related according to $R_p = 2\pi\nu_z L Q$, where $L$ is the inductance. By varying the detector $Q$ factor using electronic feedback [33] and measuring the particle dip width, as shown in Fig. 2(a), we confirm this relationship explicitly. In addition, we determine the noise temperature $T_z$ by measuring the axial energy of the single antiproton in thermal equilibrium with the detection system [34]. To perform this measurement, we couple the magnetron and axial modes, at frequencies $\nu_-$ and $\nu_z$, respectively, using a sideband drive with frequency $\nu_{\text{rf}} = \nu_z + \nu_-$ [35]. Once the drive is switched off, the probability for the antiproton to occupy a magnetron energy level with quantum number $n_-$ follows a Boltzmann probability distribution with temperature $T_- = (\nu_-/\nu_z)T_z$. In the AT, the strong magnetic inhomogeneity provided by the ferromagnetic ring electrode causes a shift in axial frequency of 40(3) $\mu$Hz when $n_-$ is changed by one quantum number. Thus, by repeatedly applying sideband drives and measuring axial frequencies, we can directly determine the Boltzmann distribution of axial energies and, hence, $T_z$. The resulting distribution of energies measured with this single antiproton thermometer is plotted in Fig. 2(b) for three different detector noise temperatures. For the settings applied to derive the ALP-to-photon conversion limits, the detector is operated at an axial temperature of $T_z = 5.7(4)$ K.

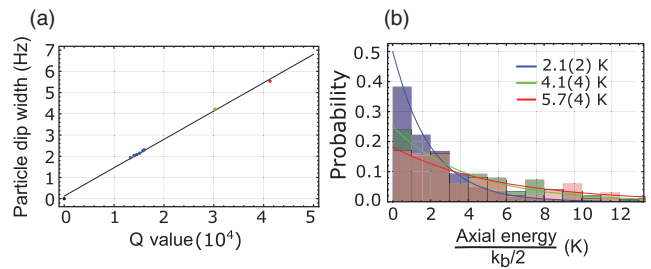

FIG. 2. (a) Explicit verification of the relationship between particle dip width and detector $Q$ factor. (b) Axial energy distribution of a single trapped antiproton used to determine the detector temperature at three different feedback settings. The measured temperatures are shown in the figure legend.

To illustrate the type of limit that can be set using the spectra of cryogenic Penning-trap resonators, we choose a sample of 1950 FFT spectra with 200 Hz span, where the particle's axial frequency was not resonant with the detector. Each spectrum was averaged over 60 s with exponential weighting with 12.5% spectrum overlap [31]. We then further linearly average many power spectra taken under identical resonator conditions yielding seven averaged spectra; we denote the averaged value of the power spectra for the $n$th averaged spectrum in the $i$th frequency bin by $\bar{d}(n)_i$. The expected value $\lambda(n)_i = \langle \bar{d}(n)_i \rangle$ is the average of the sum $V_n^2 + V_a^2$ within that frequency bin, multiplied by the amplifier gain $G$. We assume that the ALP signal has a frequency probability density function $P(\nu_a, \nu)$ equal to the dark matter background, which we model using the standard halo model [36]. The resulting expression for $\lambda(n)_i$ is

$$\lambda(n)_i = [1 + K g_{a\gamma}^2 P(\nu_a, \nu_i) S(\nu_i, n)] G V_n(\nu_i)^2, \quad (7)$$

where, assuming that the ALP signal comprises all the local dark matter density $\rho_a = 0.4$ GeV/cm$^3$ [37,38],

$$K = \left( \frac{\pi \nu_0 Q l N_T (r_2^2 - r_1^2) \|\mathbf{B}_e\|}{\sqrt{16 k_B T_z R_p}} \right)^2 \rho_a \hbar c \quad (8)$$

$$= 1.8(3) \times 10^{20} \text{ GeV}^2 \text{ s}^{-1}. \quad (9)$$

The function $S(\nu_i, n)$ captures how the strength of the signal on the detector decreases when the ALP frequency does not match the detector's resonant frequency $\nu_0$:

$$S(\nu_i, n) = \left( 1 + \frac{e_n^2}{4\kappa^2 k_B T_z R_p} f(\nu_i, Q, \mathbf{q})^{-1} \right)^{-1}. \quad (10)$$

On resonance $S(\nu_0, n) = 1.001$, and when $|\nu - \nu_0| = 100$ Hz it rises to $S(\nu, n) = 1.07$.

In order to determine whether or not we have detected an ALP signal, we follow the procedure discussed in Refs. [14,39,40]. We define the likelihood of a dataset $\bar{d}(n) = \{\bar{d}(n)_1, \ldots, \bar{d}(n)_m\}$ formed by averaging together $N$ underlying spectra, given that there is an ALP with rest mass $m_a$ and coupling constant $g_{a\gamma}$, as

$$\mathcal{L}(\bar{d}(n)|\{g_{a\gamma}, \nu_a, \mathbf{b}\}) = \prod_{i=1}^{m} \lambda(n)_i^{-N} \exp\left[ -N \frac{\bar{d}(n)_i}{\lambda(n)_i} \right], \quad (11)$$

where $\mathbf{b}$ are the background parameters associated with $V_n$. The profile log likelihood is then

$$\Theta(\nu_a, g_{a\gamma}) = 2 \ln [\mathcal{L}(\bar{d}|\{g_{a\gamma}, \nu_a, \hat{\mathbf{b}}\})] - 2 \ln [\mathcal{L}(\bar{d}|\{g_{a\gamma} = 0, \hat{\mathbf{b}}\})], \quad (12)$$

where the hats denote the values of $\mathbf{b}$ that maximize $\mathcal{L}$ for a given $g_{a\gamma}$, $\nu_a$, and $\bar{d}$. To test the data for the existence of an ALP signal, we use the test statistic $\text{TS}(\nu_a) = \Theta(\nu_a, \hat{g}_{a\gamma})$, which we correct for spectral leakage correlations between the FFT bins imposed by the windowing function. After correction, this test statistic is asymptotically $\chi^2$ distributed by Wilks' theorem, as shown explicitly in Ref. [39], Appendix D. We can therefore look for values which exceed a threshold

$$\text{TS}_T = \left[ \Phi^{-1}\left( 1 - \frac{p}{m} \right) \right]^2. \quad (13)$$

Here, $\Phi^{-1}$ is the inverse cumulative density function of a normal distribution with zero mean and standard deviation 1, $m$ is the total number of axion frequency points considered, which corrects for the look-elsewhere effect [39], and $p$ is the one-sided $p$ value for a discovery at the $n\sigma$ level. To claim a $5\sigma$ discovery for our dataset, the test statistic should exceed $\text{TS}_T = 38$. The maximum observed value is $\text{TS}_T = 9.6$, so consequently we do not observe any ALP signals in our data.

To set 95% confidence limits, we use the test statistic

$$q(\nu_a, g_{a\gamma}) = \Theta(\nu_a, g_{a\gamma}) - \Theta(\nu_a, \hat{g}_{a\gamma}) \quad \text{if } g_{a\gamma} > \hat{g}_{a\gamma}$$

$$q(\nu_a, g_{a\gamma}) = 0 \quad \text{otherwise.}$$

We search for the value of $g_{a\gamma}$ that sets $q(\nu_a, g_{a\gamma}) = -2.71$; this is the 95% confidence limit for this half-chi-square distributed test statistic after correction for spectral leakage. The limits are then power constrained following the procedure described in Ref. [41], constraining the minimum value of $g_{a\gamma}$ to be $1\sigma$ below the expected 95% confidence limit. The resulting power-constrained limits are plotted in Fig. 3(a). Here the dark blue limits are the result of the analysis conducted in this paper, and other constraints are listed in the figure caption. Notice that, in between the astronomical limits measured using the Fermi-LAT space telescope [42] (red) and limits from SN-1987A [43] (purple), there is a range of coupling strengths over which we are able to place the first limits.

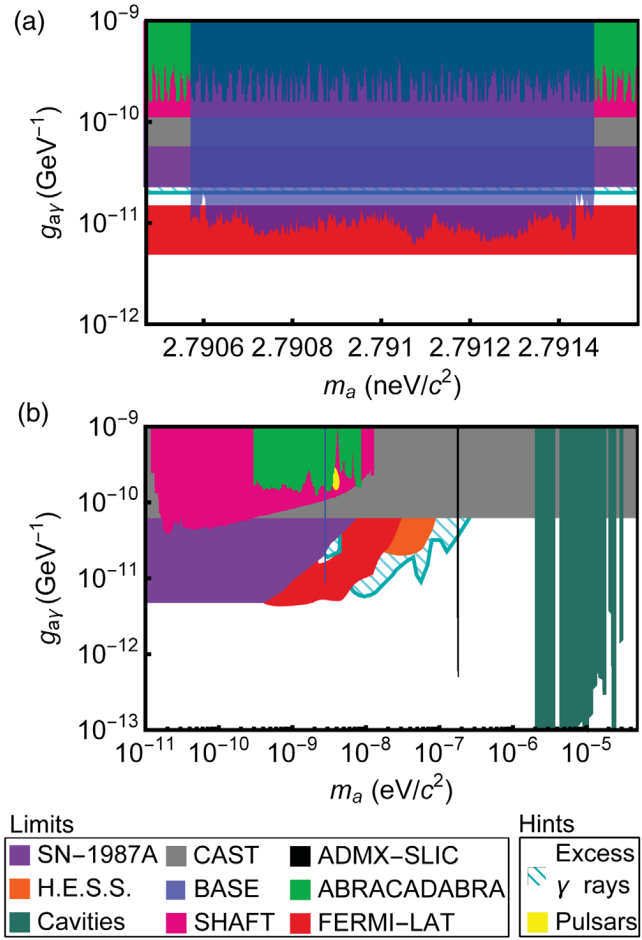

FIG. 3. (a) 95% confidence exclusion region for the coupling constant $g_{a\gamma}$ from this work shown in blue and other limits [14,19,20,42–54] and suggested regions from astronomical studies [7–9]. (b) Wider limits.

Figure 3(b) puts these limits into the wider context of other dedicated axion experiments and astronomical observations. The mass range investigated in this work is tiny compared to the other experimental approaches in this region; however, the limits achieved on $g_{a\gamma}$ are comparable to the astrophysical limits set by Fermi-LAT (red). They are around 10 times more stringent than the first-generation ABRACADABRA pathfinder experiment [14] and search for halo axions with ferromagnetic toroids (SHAFT) experiment [20] and 5 times stronger than results reported by the CAST helioscope [44]. Our limits are around a factor of 10–20 less stringent than the ADMX-SLIC experiment [19], which uses a lumped $LC$ circuit operating at 42 MHz. In part, the difference in performance can be attributed to the lower magnetic fields (2 T vs a maximum of 7 T in ADMX-SLIC) and shorter data acquisition time used in the BASE experiment. Our experiment also benefits from a direct measurement of the detector temperature using a trapped antiproton, an advantage of using the highly sensitive single-particle detectors found in Penning-trap experiments for axion and ALP searches.

To adapt these detectors into more powerful ALP search experiments with higher detection bandwidth, we are currently developing superconducting tunable capacitors. Together with a dedicated low-capacitance design of the superconducting inductor, we expect detection bandwidths in the range of 500 kHz to 1.2 MHz, at sensitivities which are at least comparable to the ones reported in this work. Placing the detector in a 7 T magnet also available at BASE and using a broader FFT span which takes advantage of the low amplifier input noise would allow the gap between 2 and 5 neV between the Fermi-LAT and SN-1987A results to be constrained to an upper limit of $1.5 \times 10^{-11}$ $\text{GeV}^{-1}$ in around one month. The detector properties could also be probed using trapped protons, which can also be loaded into the apparatus. This upgraded experiment would also investigate the favored region proposed by Ref. [9].

In this work, we have presented the first use of the ultrasensitive image current detection system of a Penning trap to search for axionlike particles with masses in the neV range. With the current setup, we place the strongest laboratory constraints for a narrow mass range around 2.791 neV, at a level which is comparable to that obtained from astrophysical observations with the Fermi-LAT space telescope and stronger than several other current haloscope and helioscope experiments. The interaction of the trapped antiproton with the detection system allows an independent determination of the detector properties, enabling the measured voltage noise signal to be straightforwardly related to the expected ALP signal. We expect that similar analyses performed on data from other Penning traps may allow comparable limits to be placed in different frequency ranges, provided the detectors in these experiments are favorably aligned. This work paves the way for future experiments which scan through broader frequency ranges at improved sensitivity.

We acknowledge technical support from the Antiproton Decelerator group, CERN's cryolab team, and all other CERN groups which provide support to antiproton decelerator experiments. We acknowledge financial support from the RIKEN Chief Scientist Program, RIKEN Pioneering Project Funding, the RIKEN JRA Program, the Max-Planck Society, the Helmholtz-Gemeinschaft, the DFG through SFB 1227 "DQ-mat," the European Research Council (ERC) under the European Union's Horizon 2020 research and innovation program (Grant Agreements No. 832848 and No. 852818) and the Max-Planck-RIKEN-PTB Center for Time, Constants and Fundamental Symmetries.